# An efficient evolutionary structural optimization method for multi-resolution designs

Hongxin Wang[1,2], Jie Liu[1*], Guilin Wen[1,2]

*[1]Center for Research on Leading Technology of Special Equipment, School of Mechanical and Electric Engineering, Guangzhou University, Guangzhou 510006, Peoples Republic of China*
*[2]State Key Laboratory of Advanced Design and Manufacturing for Vehicle Body, Hunan University, Changsha 410082, Peoples Republic of China*

*Corresponding author:
*Email:* jliu@gzhu.edu.cn

**Abstract:**

To solve large-scale or high-resolution topology optimization problem, a novel algorithm is developed based on modified bi-directional evolutionary structure optimization (BESO) and extended finite element method (XFEM). Within XFEM, a set of enriched nodes are defined to divide the finite element into several uniform sub-regions, i.e. sub-triangles and sub-tetrahedrons. The material grid and shape functions are defined on each sub-region to improve the computational accuracy, whereas the equilibrium equation is established on the level of coarse finite elements to increase the computational efficiency. We set all the standard FE nodes and the enriched nodes as the design variables, and a modified material interpolation model is introduced to calculate the material properties for sub-regions. An enrichment function originating from modeling voids scheme is adopted to character the discontinuity between solid material to void material. To efficiently use the gradient-based algorithm, BESO, sensitivity analysis is performed with the aid of adjoint method. Typical numerical examples, involving millions of design variables, are carried to verify the effectiveness of the proposed method.

## 1 Introduction:

Continuum topology optimization (TO), first developed by Bendsøe and Kikuchi [1], has been extensively studied to design lightweight, high-strength and versatile structure in various fields. However, TO traditionally plays its role in the conceptual design stage when traditional manufacturing techniques are used. More recently, it becomes possible to fabricate structure with high complex geometries with the rise of additive manufacturing (AM) technology [2]. However, despite the commercial opportunities enabled by AM, two problems that cannot be avoided are: 1) fine features of these complex structures are generally desired by the designer and 2) design problems in actual engineering are normally in large scale. Although rapidly developed these years for TO methods, it is also an uneasy task to design real engineering structures when these two problems are involved [3]. Therefore, pursuing higher computational efficiency for large scale models is the eternal objective of TO methods.

For large-scale topology optimization, the main challenge is to iteratively solve the millions of equilibrium equations during the optimization process. Normally, parallel computing is used to address the computationally intensive tasks. For the earliest work about parallel topology optimization, Borrvall and Petersson [4] used the parallel computing in combination with domain decomposition to get a high-quality resolution of realistic designs in 3D. The equilibrium equations are solved by a preconditioned conjugate gradient algorithm. Subsequently, Kim *et al.* [5] presented the parallel topology optimization to deal with large-scale structural eigenvalue-related design problems. In a recent work, Liu *et al.* [6] presented a fully parallel parameterized level set method to realize large-scale or high-resolution structural topology optimization design. In their work, the whole optimization process is parallelized, consisting of mesh generation, sensitivity analysis, assembly of the element stiffness matrices, solving of equilibrium equations, parameterization and updating of the level set function, finally the output of the computational results. Owing to the high computing capacity of Graphics Processing Units (GPUs) technology, it is also rapidly growing in popularity to accelerate topology optimization using GPU computing. It is noteworthy that the proper implementation of topology optimization using GPU computing requires a suitable technology and formulation to make good use of its potential at accelerating computing. The main contributions at this aspect mainly include the works from Wadbro et al., Martínez-Frutos et al., Ram and Sharma, *et al.* [7-9]. It is worthy to stress that reanalysis methods [10,11] are the alternative for addressing the computationally expensive topology optimization, which can predict the current structure by using the solution of previous step. For example, Amir *et al.* [12] and Long *et al.* [13] employed the reanalysis techniques to solve topology optimization problems.

The above mentioned works aim to pursue the extension of the family of feasible topology solutions, thus, often millions of degrees of freedom are involved. To reduce the computational cost, some works that purely aim to obtain the design with high-quality boundary representation using fewer degrees of freedom, are developed. Representatively, the level set method [14] has been successfully applied to an increasing variety of design problems, which allows the separation of topological description and the physical model. Fu *et al.* [15] proposed the combined methods that take full advantage of the density-based topology optimization methods, i.e. SIMP (Solid Isotropic Material with Penalization) and BESO methods, in searching the optimum solutions and the level set method in evolving smooth boundaries. Without the aid of external topological description, Wang *et al.* [16] introduced the game of building blocks into the BESO method, aiming at assembling the optimal structure using several basis elements so that obtain the smooth boundary representation. The extended finite element method (XFEM) [17] is an alternative fixed mesh approach for modeling crack or discontinuities by augmenting the shape function with an enrichment function. This method is preferred to associate with the level set description to trace the material interface [18]. In Belytschko *et al.* [19], the potential interest of the association of the XFEM and the level set description for topology optimization were identified. Subsequently, Wei *et al.* [20] combined the XFEM with the level set method to solve structural shape and topology optimization problems and generate more accurate results without increasing the mesh density and the degrees of freedom. In addition to linking with level set method, Abdi [21] implement the implicit boundary representation using isoline of strain energy to obtain the smooth and accurate representation of the design boundary for BESO method.

In this paper, we present a computational efficiently topology optimization method for large-scale models whilst maintains the high quality of boundary description. Within the framework of BESO method, the XFEM is introduced to calculate millions of nodal design variables using a coarser finite mesh. In the implementation of the presented approach, all the finite elements are divided into a set of sub-parts, such as sub-triangles for 2D planes and sub-tetrahedrons for 3D problems. A new material interpolation model is introduced to calculate the material properties and shape function on each sub-region by the nodal design variables. Following the points of Kim and Kwak [22] that "*To make the design space evolve into a better one, one may increase the number of design variables*", we define all the standard FE nodes and enriched nodes for XFEM as the design variables. Using the adjoint method, an analytical sensitivity analysis with respect to nodal design variable is developed. The novelty of the presented method is that the detailed structures with fine features could be generated using acceptable cost, especially for 3D large-scale models. And usually a personal computer is enough to deal with the problem involving millions of design variables. Moreover, the proposed triangulated partition for XFEM could improve the quality of the structural boundary representation.

The rest of the paper is organized as follows. Section 2 describes the basis theory of BESO method. Section 3 illustrates the details of XFEM. In Section 4, we outline the implementation of the proposed topology optimization algorithm. In Section 5, we present the modified sensitivity filter to avoid the appearance of material discontinuity. Several typical numerical examples are listed in Section 6 to validate the effectiveness of the proposed method. Finally, the paper is closed with some concluding remarks.

**2 Optimization formulation for BESO approach**

The implementation of the proposed method is based on the classical BESO method. Originally, the evolutionary structural optimization method (ESO) was proposed by Xie and Steven in the early 1990s [23]. The basic idea of ESO method is that optimal structural can be produced by gradually removing the ineffective material from the design domain. Later, BESO [24-26] was developed, which allows both material removal and addition. In this work, the topology optimization problem was defined to maximize structural stiffness with a volume constraint as,

$$
\begin{aligned}
&\text{Minimize} \quad && C\,(r) = \boldsymbol{F}^{\boldsymbol{T}}\boldsymbol{U} \\
&\text{subject to} \quad && V^{*} - \sum_{i=1}^{N} V_i \rho_i = 0 \\
& && \boldsymbol{KU} = \boldsymbol{F} \\
& && \rho = \rho_{\min} \ or 1
\end{aligned}
\tag{1}
$$

where $\boldsymbol{F}$ is the global load vector, $\boldsymbol{U}$ is the global displacement vector. $V_i$ and $\rho_i$ are respectively the density and volume of an individual element, $V^{*}$ denotes the prescribed structural volume. In order to avoid numerical difficulties associated with zero density elements, a small value of $\rho_{\min}$ is used to denote the void material. $\boldsymbol{K}$ is the global stiffness matrix, which is constructed by:

$$\boldsymbol{K}=\sum_{i=1}^{N} k_e(E_e) \tag{2}$$

where $k_e$ is the stiffness matrix of basic element, $N$ is the total number of all the elements. The relation between the Young modulus and the material density is defined as,

$$E_{\mathrm{e}}=\rho_e^p E_0 \tag{3}$$

where $E_0$ is the Young's modulus of solid material. The penalty exponent $p$ is typically chosen as $p$=3.

## 3 Extended finite element method

As well known, the XFEM possesses the potential to associate with the topology optimization methods to obtain the smooth boundary representation. When the design boundary cuts across the element edge, the finite element is partitioned into several sub-parts, then, the element stiffness matrix is the sum of the solid material parts and weak material part. In the presented method, we partition all the elements into several sub-parts, and attempt to independently describe the material properties and shape function for each sub-part so that generates more structural details.

### *3.1 X-FEM approximation*

The conventional finite element method is troublesome at handling the moving discontinuities model analysis, due to the mesh needs to match the geometry of discontinuity. The XFEM is developed to capture the non-smooth displacement fields across material interfaces by extending the shape function with an augmented part. For different cases, such as the hole and inclusion problem [27] and the crack problem [28], several different approximation schemes have been proposed. In the hole and inclusion problems, the shape function of the conventional finite element is extended to the following form:

$$u^e(x)=\sum_{i=I} N_i(x)u_i+\sum_{i=I^*} N_i(x)\varphi_i a_i \tag{4}$$

On the right hand of Eq. (4), the first item denotes the displacement field of the standard FE part, and the second item is the displacement approximation for the enriched nodes. Where $N_i(x)$ are the conventional shape functions associated to the nodal degrees of freedom. $u_i$ and $a_i$ respectively denotes the nodal coefficients of the standard FE and enriched part. $\varphi_i$ is the problem-dependent enrichment function.

Loehnert et al. [29] pointed out that '*The major drawback of the XFEM was the fact that in the blending elements constituting the part of the mesh between the fully enriched domain to the non-enriched domain, the partition of unity is not fulfilled. This drawback does not exist in the partition of unity method (PUM) [30], the XFEM is based on.*' Therefore, in the presented approach, all the nodes within the mesh are enriched with the same set of enrichment function.

*3.2 Precondition*

In Fig. 1, the scheme of triangulated partition for XFEM is shown to illustrate the division of one four-node finite element into a set of sub-triangles. Parameter *En* denotes the number of enriched nodes along one element side. For simplicity's sake, we use the parameter *En* to represent the scheme of triangulated partition in the following. The black nodes are the standard FE nodes, and the white nodes are the enriched nodes. For each sub-triangle, there are two options of Gauss points, one is the midline Gauss points and the other is the inside Gauss points. The Gauss weight for each Gauss point is the same and equals to 1/3. In this paper, the first option of Gauss points is chosen for 2D problem.

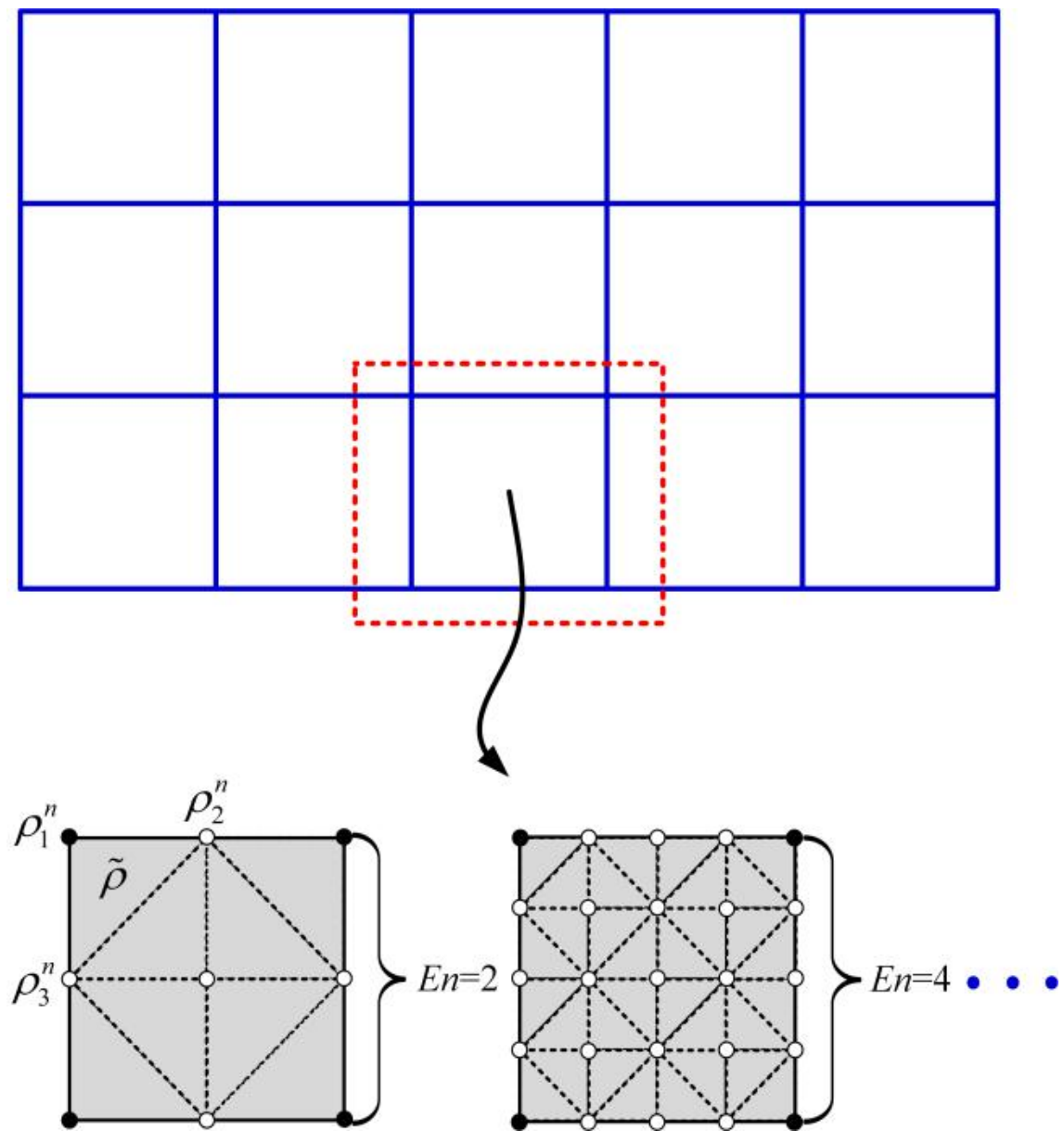

**Fig. 1** The scheme of triangulated partition using different numbers of enriched nodes.

The stiffness matrix of the solid sub-triangle is formulated as,

$$\tilde{k}_e^i = \iint_{\Omega_e} f(x,y)dxdy = A_i \sum_{j=1}^{n} w_j f(\omega_1^j, \omega_2^j, \omega_3^j) \quad (5)$$

where $f = B^T DBt$ and $i$ denotes the $i$th sub-triangle in one element. $D$ is the elastic matrix and $B$ is the strain matrix relating the displacement and the strain. $n$ is the number of gauss points, and $\omega$ is the natural coordinates of the gauss points. $A_i$ is the area of the $i$th triangle and $w_i$ is the weighting factor. Assembling the stiffness of all the sub-triangles into the element stiffness matrix and becomes,

$$\bar{k}_e = \sum_{i=1}^{N} \sum_{j=1}^{n} A_i w_j f(\omega_1^j, \omega_2^j, \omega_3^j) = \sum_{i=1}^{N} \tilde{k}_e^i \quad (6)$$

where $N$ is the number of sub-triangles for one element.

It is easy to extend the 2D XFEM to 3D case. In Fig. 2, the decomposition of one hexahedron into sub-tetrahedrons is illustrated. According to the numbers of enriched nodes, one hexahedron is divided into several sub-hexahedrons, and each sub-hexahedron is partitioned into 5 sub-tetrahedrons. The classical Gauss quadrature method is adopted to calculate the integrals over the sub-tetrahedrons. Compared to the normal hexahedron element that integrated with only 8 gauss points, the proposed XFEM integration scheme requires a higher number of gauss points.

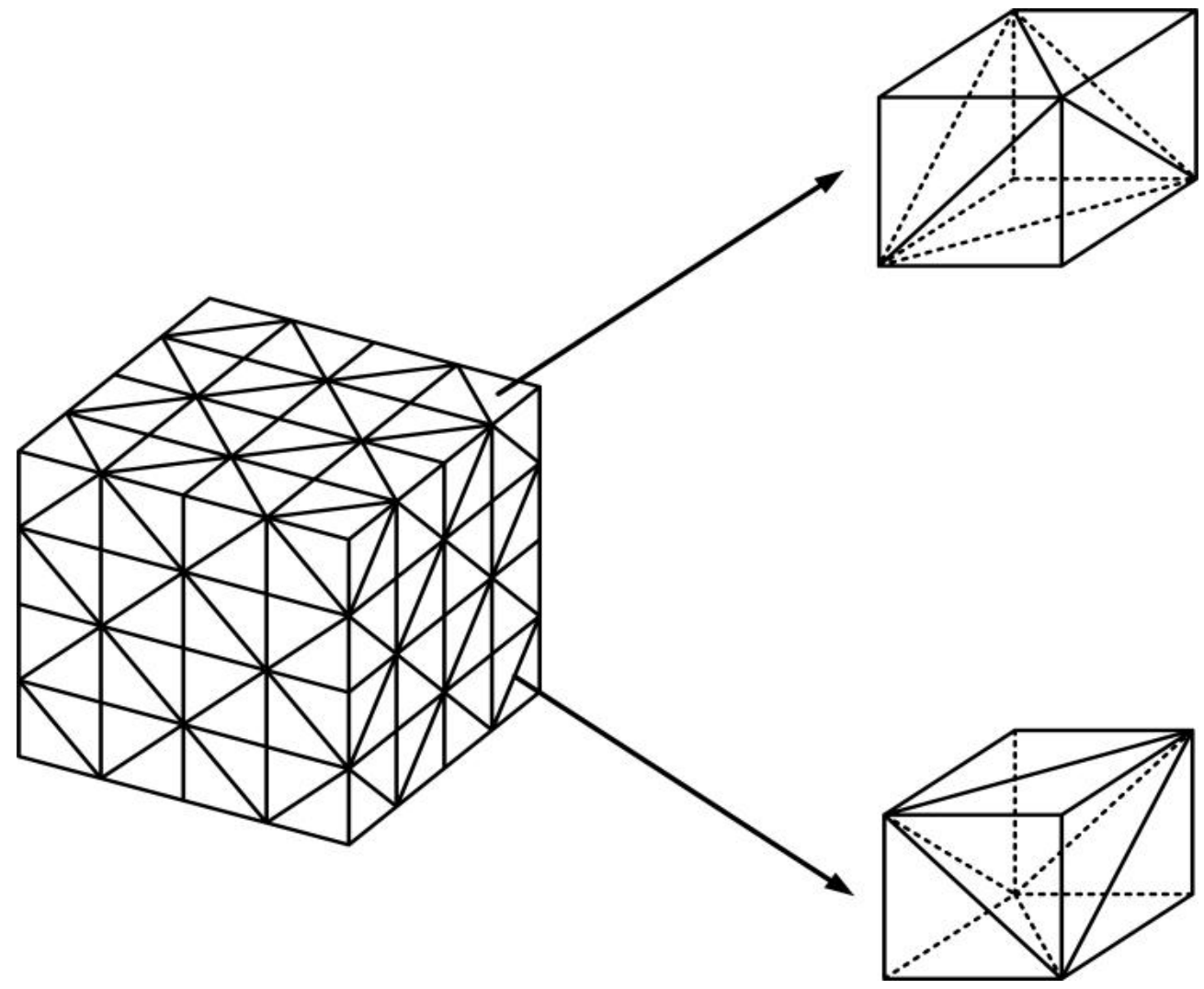

**Fig. 2** Triangulated partition for one hexahedral finite element.

## 4 Implementation details of the proposed approach

In the proposed approach, all the standard FE nodes and the enriched nodes are set as the design variables. The sub-parts partitioned from finite elements are endowed with individual material properties and shape functions. Assembling the stiffness matrix of all the sub-parts, the global stiffness matrix keeping fewer degrees of freedom is obtained. Then, an efficient sensitivity analysis procedure is presented to calculate the sensitivity information for all the nodal design variables. Next, the implementation of the proposed approach for 2D problem is illustrated.

*4.1 Material interpolation model*

The binary design variable $\rho_m^n$ is adopted to declare the absence or presence of one node. The density which does not carry much physical meaning is endowed to each sub-triangle, and is calculated by the connected nodes (as illustrated in Fig. 1),

$$\tilde{\rho}_i = \frac{1}{3}\sum_{m=1}^{3}\rho_m^n, \quad \rho_m^n = \rho_{\min}\ or\ 1 \tag{7}$$

It is easily understood from Eq. (7) that the sub-triangle with intermediate density must be occurred to the interface between the solid and void materials. In order to remove the transition regions, the intermediate densities are projected by a Heaviside function to give zero density for $\tilde{\rho}_i < 1$ and only density one for $\tilde{\rho}_i = 1$.

$$\overline{\rho}_i = e^{-\beta(1-\tilde{\rho}_i)} - (1-\tilde{\rho}_i)e^{-\beta} \tag{8}$$

By experience, the parameter $\beta$ is set to be 100. Knowing the density of each sub-triangle, the material Young's modulus without penalization is interpolated:

$$\overline{E}_i = \overline{\rho}_i E_0 \tag{9}$$

*4.2 The optimization formulation*

In this paper, the enrichment function [40] originates from modeling voids scheme is referred, and a Heaviside function is used to represent the material void discontinuity between sub-triangles [43],

$$u^e(x) = \sum_{i=I} N_i(x)H(x)u_i \tag{10}$$

where the Heaviside function is equal to 1 for the sub-triangle in the solid material and switches to 0 for the sub-triangle in the void material. Since the density of the sub-triangle is projected by the Heaviside function, the extended shape function could also be expressed,

$$u^e(x) = \sum_{i=I} N_i(x)\overline{\rho}_i u_i \tag{11}$$

Substituting Eq. (9) and Eq. (11) into Eq. (5), the stiffness matrix of the sub-triangle becomes,

$$\tilde{k}_i = A_i \sum_{j=1}^{n} w_j B^T(\overline{\rho}_i) D(\overline{\rho}_i) B(\overline{\rho}_i) t = \overline{\rho}_i^3 \tilde{k}_0(E_0) \tag{12}$$

where $\tilde{k}_0$ is the stiffness matrix of sub-triangle with solid material. Summing the contributions of all the sub-triangles, the stiffness matrix for each finite element becomes,

$$\overline{k}_e == \sum_{i=1}^{N} \overline{\rho}_i^3 \tilde{k}_0(E_0) \tag{13}$$

Assembling the local stiffness matrix into global stiffness matrix $\overline{\boldsymbol{K}}$,

$$\overline{\boldsymbol{K}} = \sum_{e=1}^{Tol} \sum_{i=1}^{N} \overline{\rho}_i^3 \tilde{k}_0 \tag{14}$$

As analysis above, the compliance minimization problem for the proposed method is formulated as:

$$\begin{aligned}
&\text{Minimize} \quad C\,(r_m^n) = \boldsymbol{F}^{\boldsymbol{T}}\boldsymbol{U} \\
&\text{Subject to} \quad \frac{\sum_{e=1}^{Tol}\sum_{i=1}^{N} \overline{V}_i \overline{\rho}_i}{V_0} \le f \\
&\qquad \overline{\boldsymbol{K}}\boldsymbol{U} = \boldsymbol{F} \\
&\qquad \rho_m^n = \rho_{\min}\ or\, 1
\end{aligned} \tag{15}$$

where $\overline{V}_i$ is the volume for each sub-triangle. For simplicity's sake, we call the proposed method as X-BESO method in the following sections of this work.

*4.3 Sensitivity analysis*

In this case, the sensitivity analysis of the objective function is performed via adjoint method. Using the chain rule, the derivative of the objective function with respect to nodal design variables is,

$$\frac{\partial C}{\partial \rho_m^n} = \frac{1}{M} \sum_{i}^{M} \frac{\partial C}{\partial \overline{\rho}_i} \frac{\partial \overline{\rho}_i}{\partial \tilde{\rho}_i} \frac{\partial \tilde{\rho}_i}{\partial \rho_m^n} \tag{16}$$

where M is the total number of sub-triangles connected to the *m*th node. The derivate of the objective function with respect to the density of sub-triangle is writhen,

$$\begin{aligned}
\frac{\partial C}{\partial \overline{\rho}_i} &= \boldsymbol{F}^T \frac{\partial \boldsymbol{U}}{\partial \overline{\rho}_i} - v\left(\frac{\partial \overline{\boldsymbol{K}}}{\partial \overline{\rho}_i}\boldsymbol{U} - \overline{\boldsymbol{K}}\frac{\partial \boldsymbol{U}}{\partial \overline{\rho}_i}\right) \\
&= (\boldsymbol{F}^T - v\overline{\boldsymbol{K}})\frac{\partial \boldsymbol{U}}{\partial \overline{\rho}_i} - v\frac{\partial \overline{\boldsymbol{K}}}{\partial \overline{\rho}_i}\boldsymbol{U}
\end{aligned} \tag{17}$$

The adjoint vector $v$ is chosen according $\boldsymbol{F}^T\text{-}v\overline{\boldsymbol{K}}=0$ , to eliminate the displacement sensitivity,

$$\frac{\partial C}{\partial \overline{\rho}_i} = -\boldsymbol{U}^T \frac{\partial \overline{\boldsymbol{K}}}{\partial \overline{\rho}_i} \boldsymbol{U} \tag{18}$$

Substituting the global stiffness matrix of Eq. (14) into Eq. (18), yields,

$$\frac{\partial C}{\partial \overline{\rho}_i} = -u_e^T \frac{\partial \tilde{k}_i}{\partial \overline{\rho}_i} u_e \tag{19}$$

where $u_e$ is the displacement field of the element to which the sub-triangle belongs, and $\tilde{k}_i$ denotes the stiffness matrix of the sub-triangle. The derivate of sub-triangle density with respect to nodal design variable is easily obtained as,

$$\frac{\partial \tilde{\rho}_i}{\partial \rho_m^n} = \frac{1}{3} \tag{20}$$

*4.4 Basic procedure*

The optimization procedure of the X-BESO approach is briefly summarized as follow:

1. Discretize the design domain using the finite mesh and define the loading and boundary condition.

2. Define a set of enriched nodes and divide the element into several sub-regions. Calculate the stiffness matrix and the volume for each sub-region.

3. Assemble the global stiffness matrix as section 4.2 and perform FE analysis to obtain the nodal sensitivity numbers.

4. Average the sensitivity number with its history information as Eq. (21) and save the resulting sensitivity number for next iteration.

$$\alpha_i = \frac{\alpha_i^k + \alpha_i^{k-1}}{2} \tag{21}$$

5. Determine the target volume for the next iteration as the regulation $V_{k+1} = V_k(1 \pm ER)$ , where $ER$ is called the evolutionary volume ratio. Once the objective volume is reached, the volume $V_{k+1}$ will be kept constant for the remaining iteration.

6. Update the nodal density. Define the threshold of sensitivity numbers for removing and adding nodes $\alpha_{del}^{th}$ and $\alpha_{add}^{th}$ . For solid nodes, it's nodal density will be switched to 0 if $\alpha_i \le \alpha_{del}^{th}$ . On the contrary, the nodal density will be switched to 1 if $\alpha_i > \alpha_{add}^{th}$ for void nodes. Project the gray sub-regions into solid or void material by Heaviside function.

7. Go back to step 3 until the objective volume is reached and the convergence criterion Eq. (22) is satisfied.

$$change = \frac{\left|\sum_{i=1}^{M}(C_{k-i+1} - C_{k-M-i+1})\right|}{\sum_{i=1}^{M} C_{k-i+1}} \leq \tau \tag{22}$$

where $\tau$ is the teeny allowable convergence criterion, and $M$ is the selected iteration number that are expected with stable compliance.

## 5 Filter scheme

As we know, the density-based topology optimization method is prone to the pathologies, namely the checkerboard and mesh-dependency, if no regularization scheme is applied. In this section, a MBB beam is optimized to investigate whether the X-BESO method itself has the function at solving these two problems. The load and boundary conditions of the MBB are illustrated in Fig. 3. The available material will cover 50% of the design domain. The other parameters are: $E$=1 Mpa, $ER$=1%, passion ratio $\mu$ =0.3. Fig. 4 gives a series of optimum designs that not consider the filter scheme. By observations, we can find that the checkerboard pattern is better improved but the optimization solutions are still mesh-dependent. Moreover, a new problem of material discontinuity is occurred, especially when utilizing coarser mesh and larger enriched nodes. In order to overcome these problems, a suitable filter scheme, such as the sensitivity filter, is still needed for the X-BESO method. Observing from Fig. 4, we can find that the proposed triangulated partition dramatically smoothes the staggered edges, and the completely smoothed edges could be produced (circled in the red).

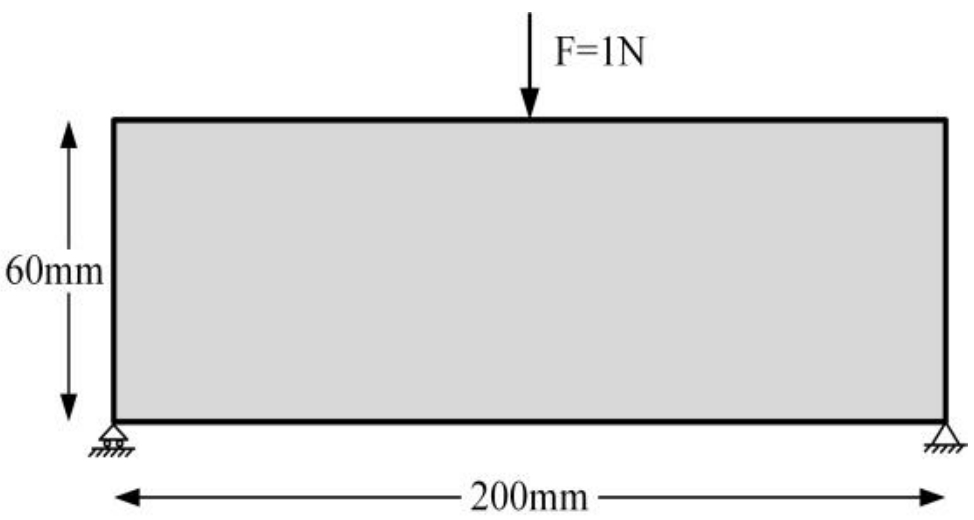

**Fig. 3** Design domain and the boundary conditions for MBB beam.

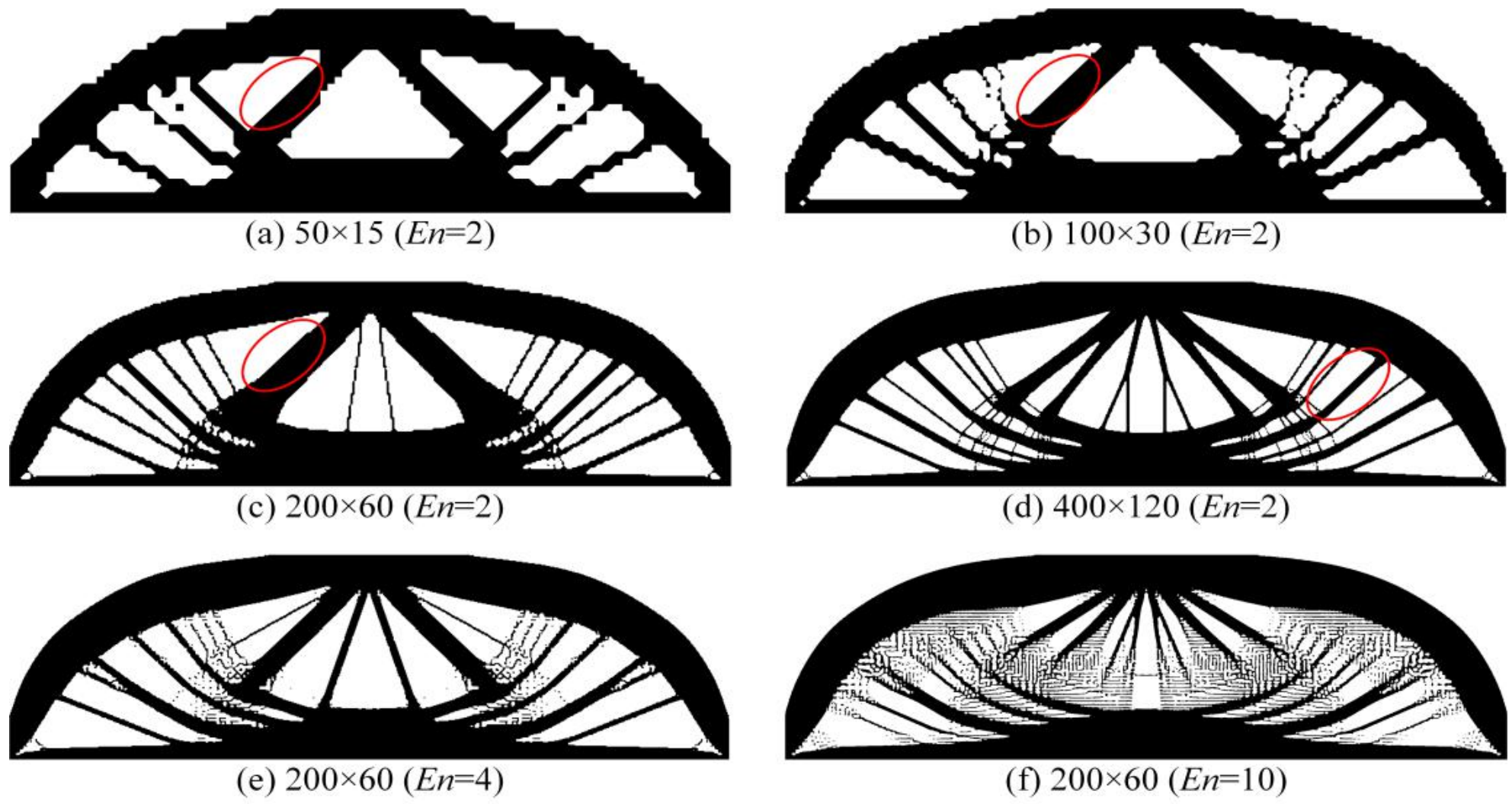

**Fig. 4** Illustration of checkerboard and mesh-independency solutions for the X-BESO method without filter scheme.

*5.1 Formulation of sensitivity filter*

The technology of sensitivity filter [31] is to convert the nodal sensitivities into smoothed sensitivity numbers. In BESO method, the standard sensitivity filter is formulated as,

$$dc = \frac{1}{\sum_{j=1}^{N} H_j} \sum_{j=1}^{N} H_j \frac{\partial c}{\partial \rho_m^n} \tag{23}$$

where $N$ denotes the total number of nodes that locates in the filter radius $r_{\min}$. The convolution operator (weight factor) $H_j$ is written as,

$$H_j = r_{\min} - dist(m, j) \tag{24}$$

$$\left\{ j \in N \mid \mathrm{dist}(e, j) \le r_{\min} \right\}, \quad e = 1, \ldots N$$

where the operator $\mathrm{dist}(m, j)$ is defined as the distance between the center of node $m$ and node $j$. The convolution operator $H_j$ equals to zero outside the filtering area and decays linearly with the distance from node $m$ to node $j$. Fig. 5 gives an illustration of the sensitivity filter for the X-BESO method. By drawing a circle of radius $r$ centered at $m$th node, the nodes that will influence the sensitivity of $m$th node are identified.

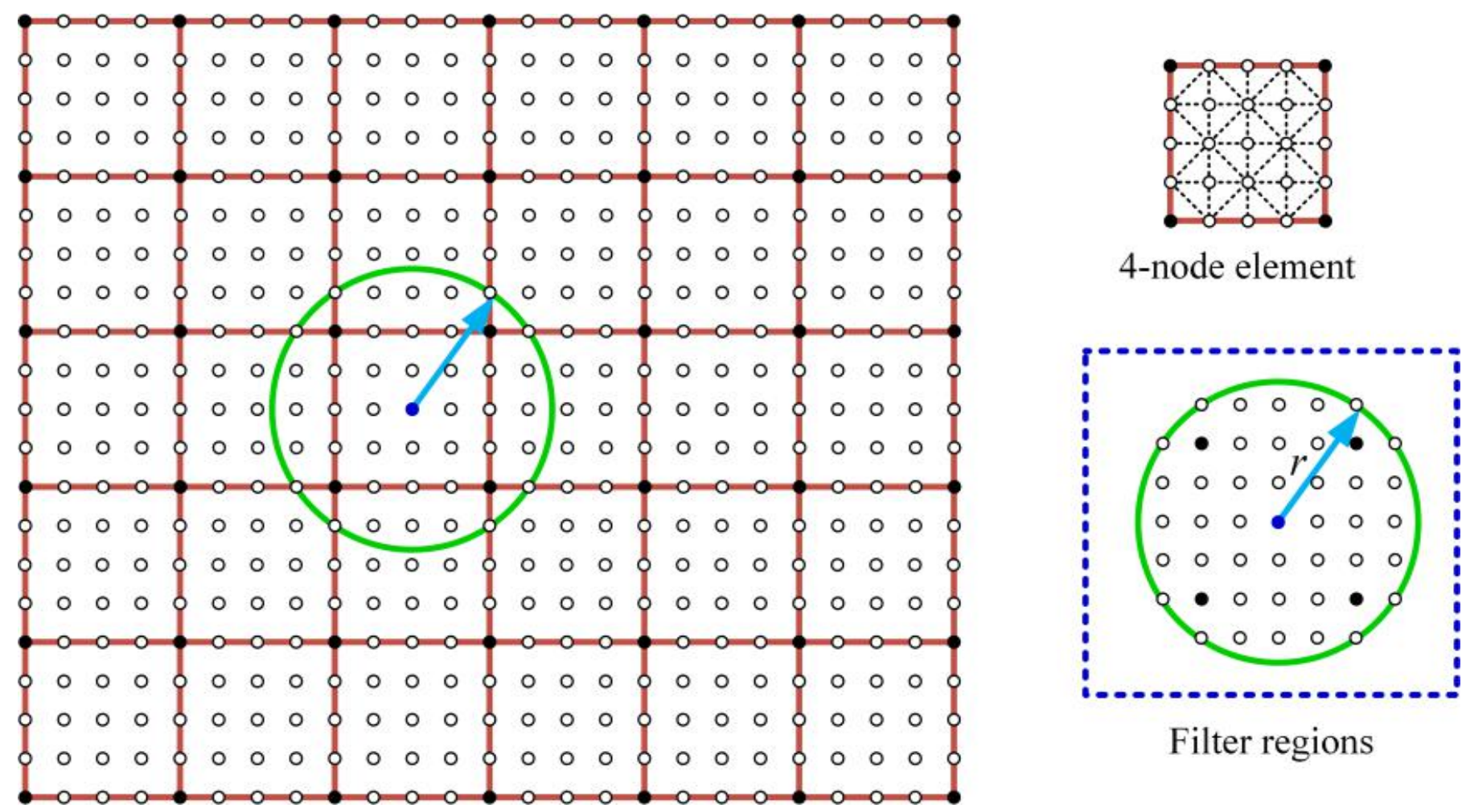

**Fig. 5** Illustration of the sensitivity filter in the X-BESO method.

In order to study the applicability of the sensitivity filter in the X-BESO method, the above MBB problem is optimized on a finite mesh 200×60 by using the X-BESO method. The enriched nodes *En*= 10 are defined for each finite element, thus, the allocation of the nodal design variables is 2001×601.Three different filter radii r=6, 8 and 30 are respectively tested. The final topologies are shown in Fig. 7(a)-(c). It can be found that the phenomenon of material discontinuity exists if the filter radius is smaller than the width of one element. To solve this problem, a modified sensitivity filter with non-linear convolution operator is introduced in this paper.

*5.2 Modified sensitivity filter*

In original sensitivity filter, the convolution operator (weight factor) decays linearly with the distance from central node to surrounding nodes. But we found that the steep weighting gradient easily causes the poor connection between central node to surrounding nodes, leading to material discontinuity and one-node connected hinge. To this end, we introduce a new convolution operator with non-linear weighting gradient,

$$H_f^{'} = e^{-\lambda dist(m,j)/r_{\min}} \tag{25}$$

$$\left\{ j \in N \mid \mathrm{dist}(\mathrm{m}, j) \le r_{\min} \right\}, \quad m = 1, \ldots N$$

where the definition of $\mathrm{dist}(m,j)$ is the same as that of original convolution operator. The convolution operator $H_f^{'}$ decays non-linearly with the distance from node *j* to node *m*.

Identically, $H_f^{'}$ equals to zero for the node outside the filter area. The coefficient $\lambda$ is to control the non-linear degree of the weighting gradient. Fig. 6 plots the gradient curves of the non-linear convolution operator for different coefficients $\lambda$. It can be observed that the smaller the parameter $\lambda$ takes, the flatter the weighting gradient is.

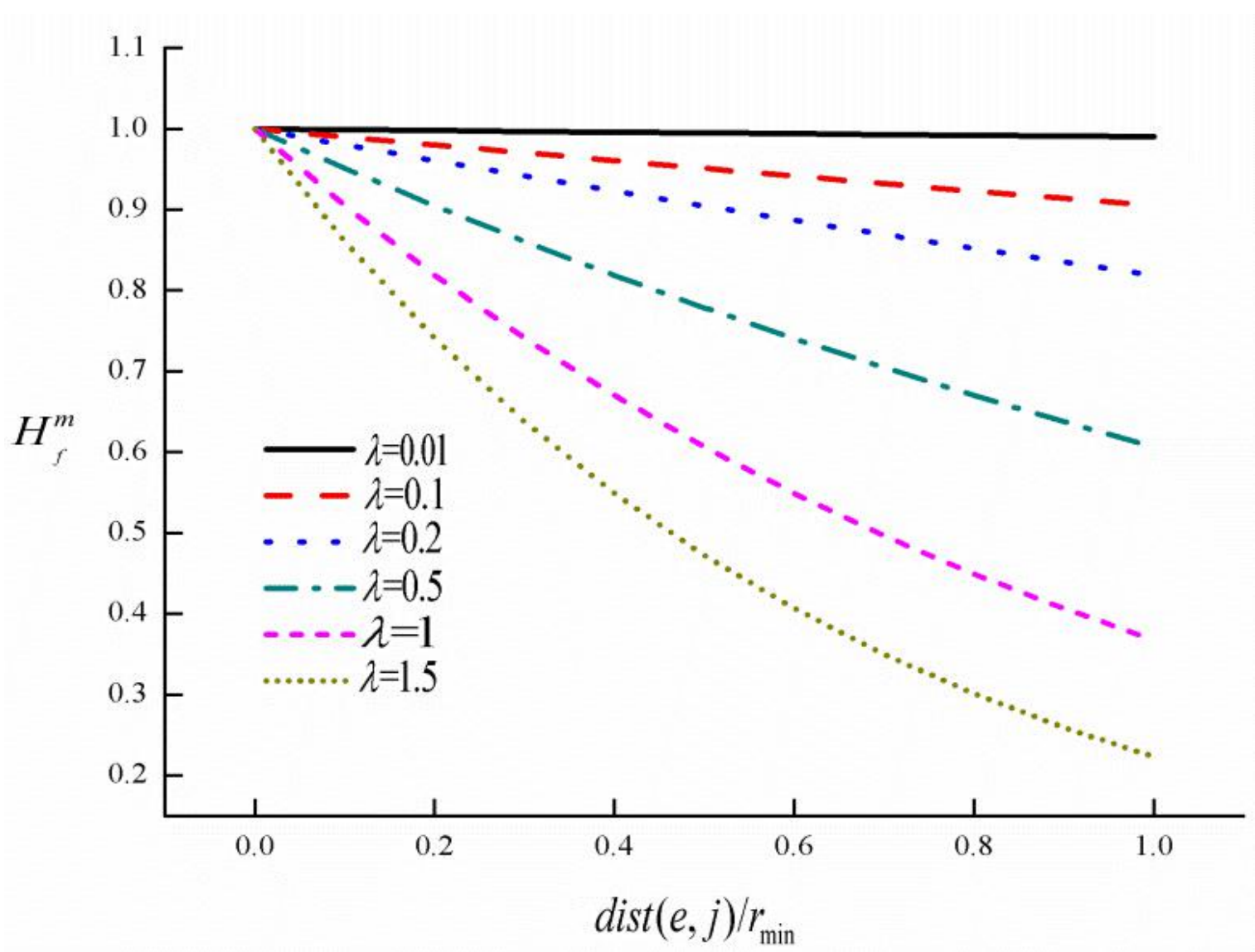

**Fig. 6** Gradient curves of the non-linear convolution operator for different coefficients $\lambda$.

To test the non-linear convolution operator, the X-BESO method adopting the modified sensitivity filter with five schemes of parameters, namely $\lambda$=0.01, 0.1, 0.2, 0.5 and 1 is tested on the MBB beam. The case of discretizing the domain with coarser mesh 200×60 coupled with larger enriched nodes *En*=10 is studied, and the final designs are given in Figs. 7(d)-(j). Apparently, the modified sensitivity filter greatly improves the problem of material discontinuity (circled in the red). For filter radius *r*=8, the phenomenon of material discontinuity are absolutely suppressed for $\lambda \le 0.5$. But for smaller filter radius *r*=6, the phenomenon of material discontinuity may still exist. For larger filter radius, such as *r*=30, the modified sensitivity filter removes some tiny members and achieve the minimum length scale control on solid material. As analysis above, the parameter $\lambda$=0.2 is adopted in the following examples.

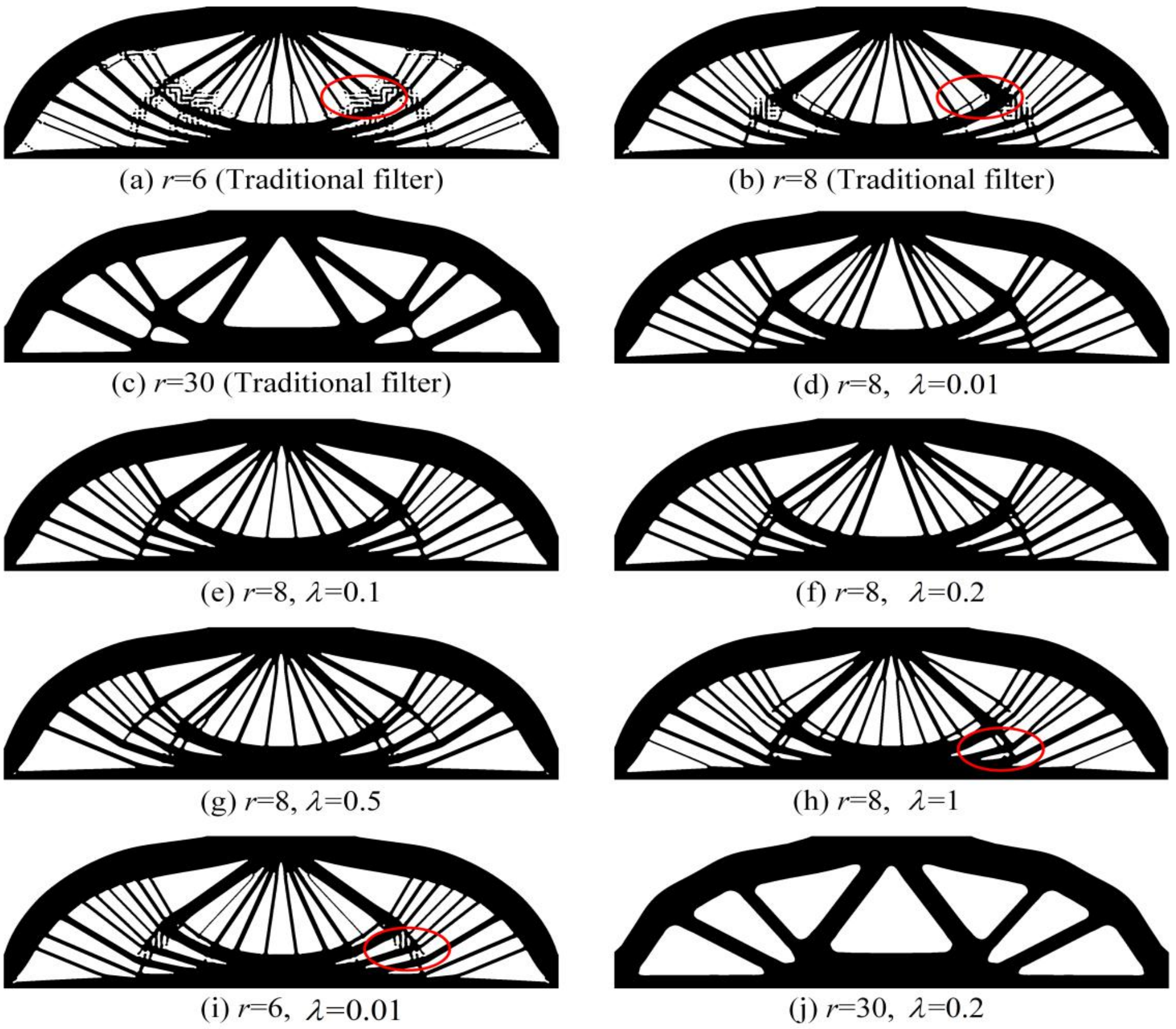

**Fig. 7** The X-BESO designs obtained from: (a)-(c) traditional sensitivity filter; (d)-(j) the modified sensitivity filter.

## 6 Results and discussions

To illustrate the capability of the X-BESO approach, we consider here a series of classical minimum compliance problems for 2D and 3D structures. In all subsequent examples, the solid material is uniformly assumed with Young's modulus *E*=1 Mpa and Poisson's ratio 0.3, if without special declaration. The unit mm is adopted in the paper. The evolutionary volume ratio *ER*=1% is used for 2D problem, and *ER*=2% for 3D structures. All experiments were conducted on a personal office computer equipped with Intel Xeon CPU E5-2620, 8 cores and 16 GB memory. It should be emphasized that just one core of the computer is used.

### *6.1 Cantilever beam*

A cantilever structure with unit thickness is optimized in this example. The geometrical dimension of the domain is shown in Fig. 8. The left side of the domain is fixed, and a unit external load $\boldsymbol{F}$=1N is applied downward at the center of the free. The volume constraint is 50% of the design domain.

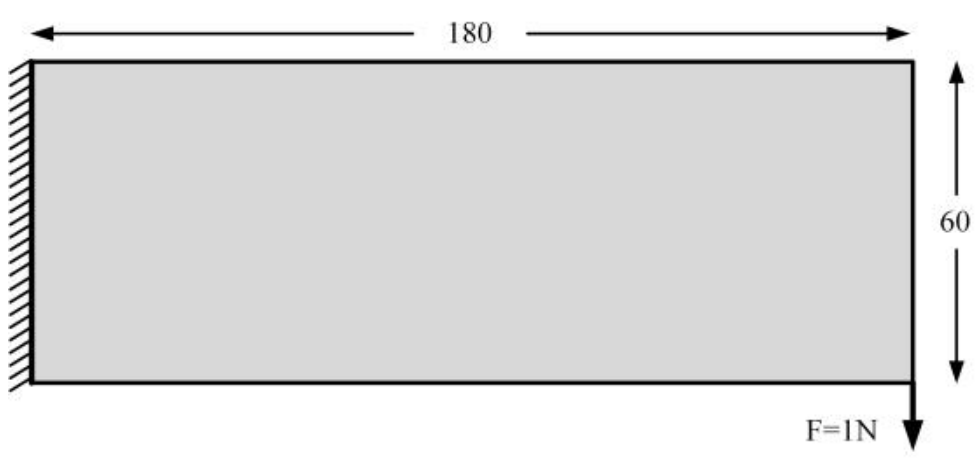

**Fig. 8** Design domain and boundary conditions for the cantilever beam.

Fig. 9 respectively shows the final designs optimized by different methods, different meshes and different filter schemes. In the figures, *C* is the compliance with a unit of N.mm. *t* denotes the CPU time consumed at each step, and r is the filter radius. By comparison, the X-BESO method generates the results possessing similar structures as traditional BESO designs, which illustrates the accuracy of the X-BESO method. But it should be noted that the X-BESO designs show better mechanical performances in terms of the stiffness of the structure. In the right of Fig. 9, some members thinner than the filter radius are removed, illustrating that the modified sensitivity filter perform better at minimum length scale control on solid material. In terms of the computational efficiency, the X-BESO approach saves more than 6 times CPU time than the BESO method to generate the same-resolution structures, when utilizing coarser finite mesh 180×60 coupled with larger enriched nodes *En*=10. In one word, the computational efficiency is greatly improved when using smaller-scale equilibrium equations. Fig. 10 plots the evolution history of the mean compliance and the volume fraction to prove the good convergence of the presented approach.

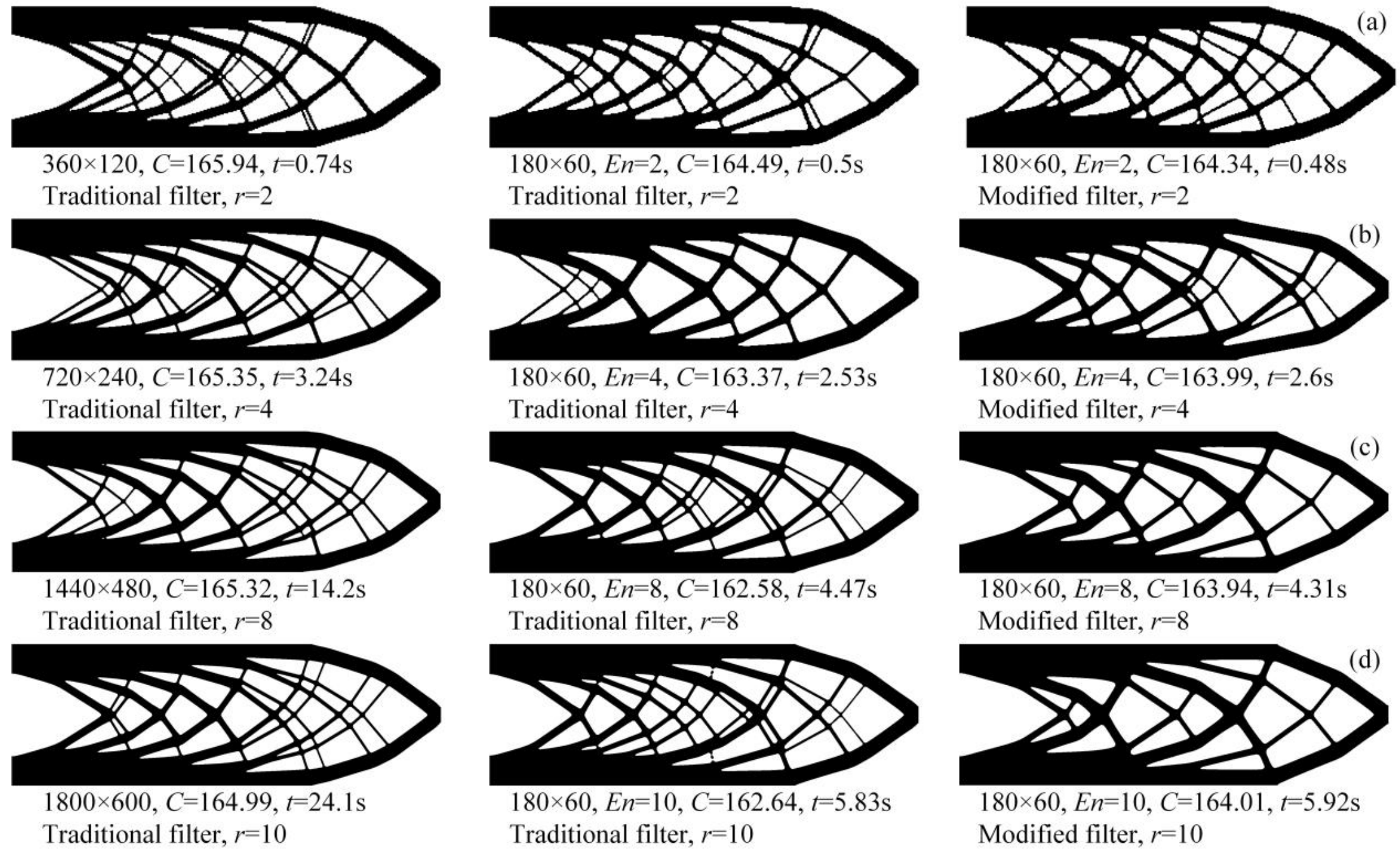

**Fig. 9** The optimum designs obtained from different methods, different meshes and different filter radii.

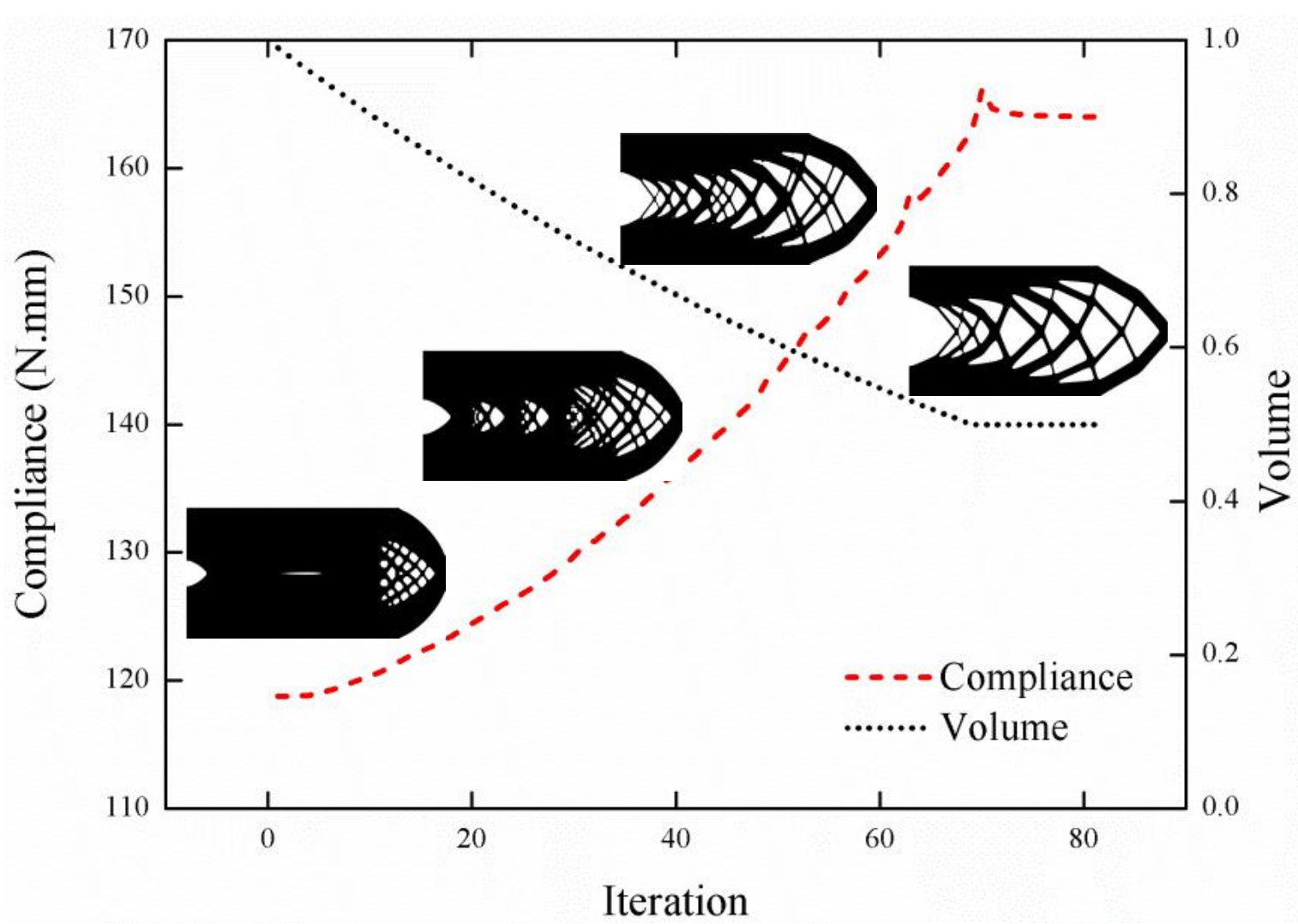

**Fig. 10** Evolution histories of mean compliance and volume fraction for the X-BESO method.

*6.2 L-brackets*

This example optimizes the L-bracket with different filter radii, aiming to investigate the effect of filter radius on the X-BESO approach. The top end of the L-bracket is clamped and the external load ***F***=1 N is imposed on the top of the lower right corner (shown in Fig. 10).We assume that the available material can cover 30% of the design domain.

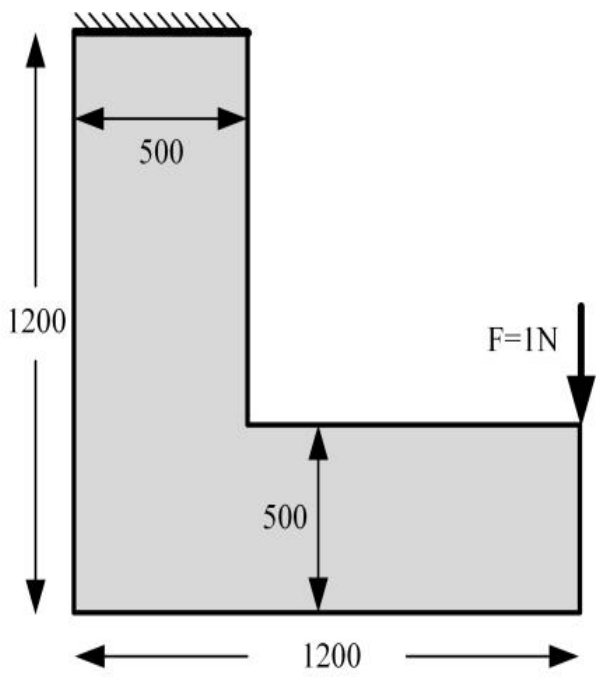

**Fig. 11** Design domain and boundary conditions for the L bracket.

Fig. 12 gives the topology designs obtained from different methods and different filter radii. For BESO method, the finite mesh 1200×1200 is used. For X-BESO method, the background finite mesh is 120×120, and each element is partitioned using enriched nodes *En*=10. Apparently, the effectiveness of the modified sensitivity filter at suppressing material discontinuity can be observed for the case of *r*=8. Using different filter radii, the structures with different details are yielded from the X-BESO method. In addition, the X-BESO designs also perform stiffer than the standard BESO designs. Similar to previous case, the X-BESO approach saves nearly 6 times computational cost than the BESO method.

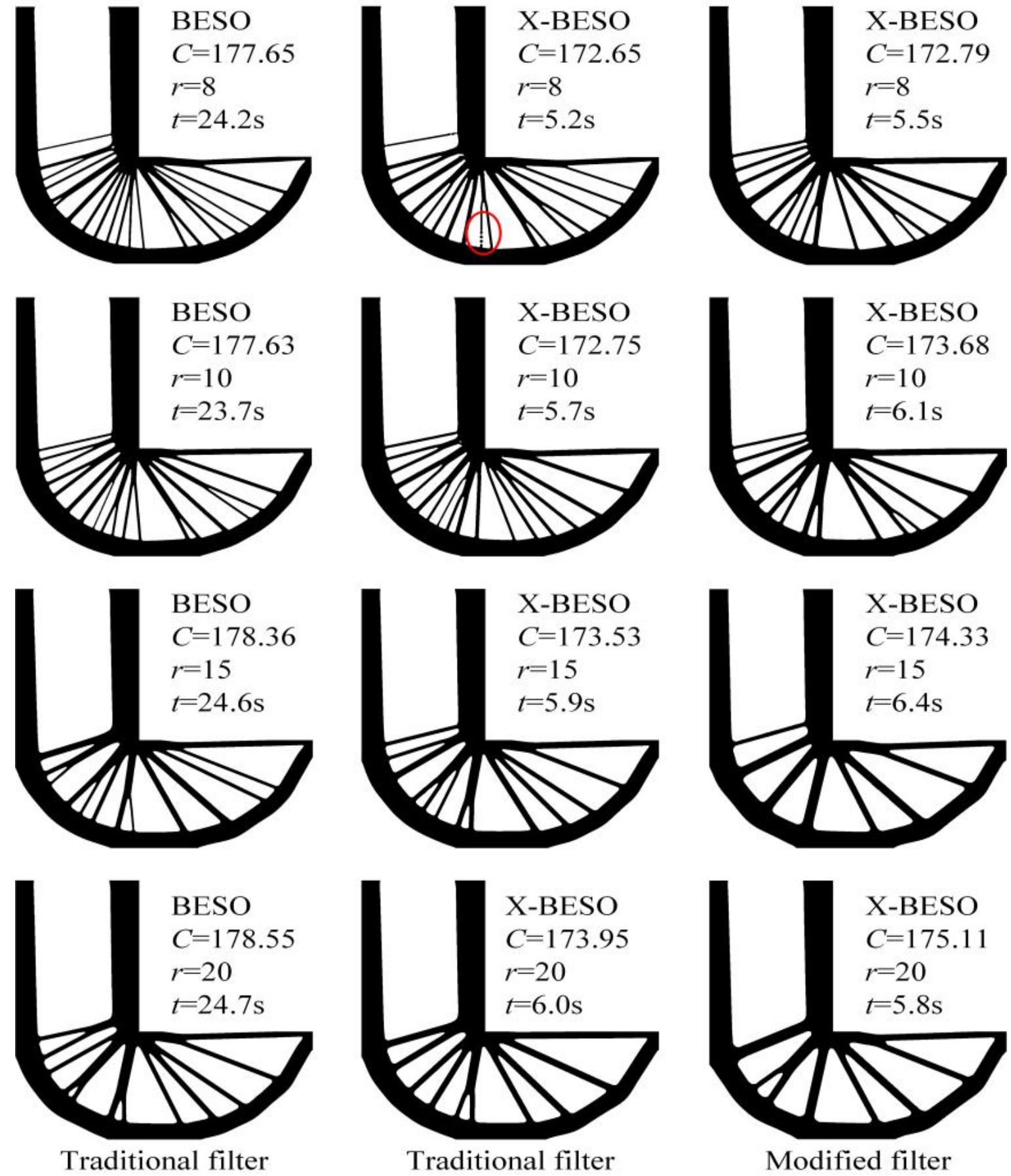

**Fig. 12** The optimum designs obtained from different methods and different filter radii.

*6.3 3D cantilever beam*

In this case, a three-dimensional cantilever beam model is optimized. Since the limitation of computing resource, just the optimization results obtained from the X-BESO method are shown in this paper. The left end of the cantilever beam is fixed in all three directions, and a vertical downward concentration force is applied in the center of the right end (as shown in Fig. 13(a)).The background finite mesh is 60×20×10, and the enriched nodes $En$=8. The 481×161×81 nodal design variables are uniformly allocated in the domain. The maximum usable volume fraction is defined as 0.12.

Figs. 13(b)-(c) show the topology solutions optimized with filter radii r=8, 12. Note that all the following 3D examples are optimized with the traditional sensitivity filter. The total number of nodal design variables is 6272721. For these two topology solutions, respectively 13h and 13.3h

are used. From the comparison, we can find that the optimized result possesses more truss-like components when using the smaller filter radius. For the case of utilizing the larger filter radius, some thinner components are removed but the main structural members are strengthened. It is noteworthy that the X-BESO alleviates the requirement on the computer's memory space.

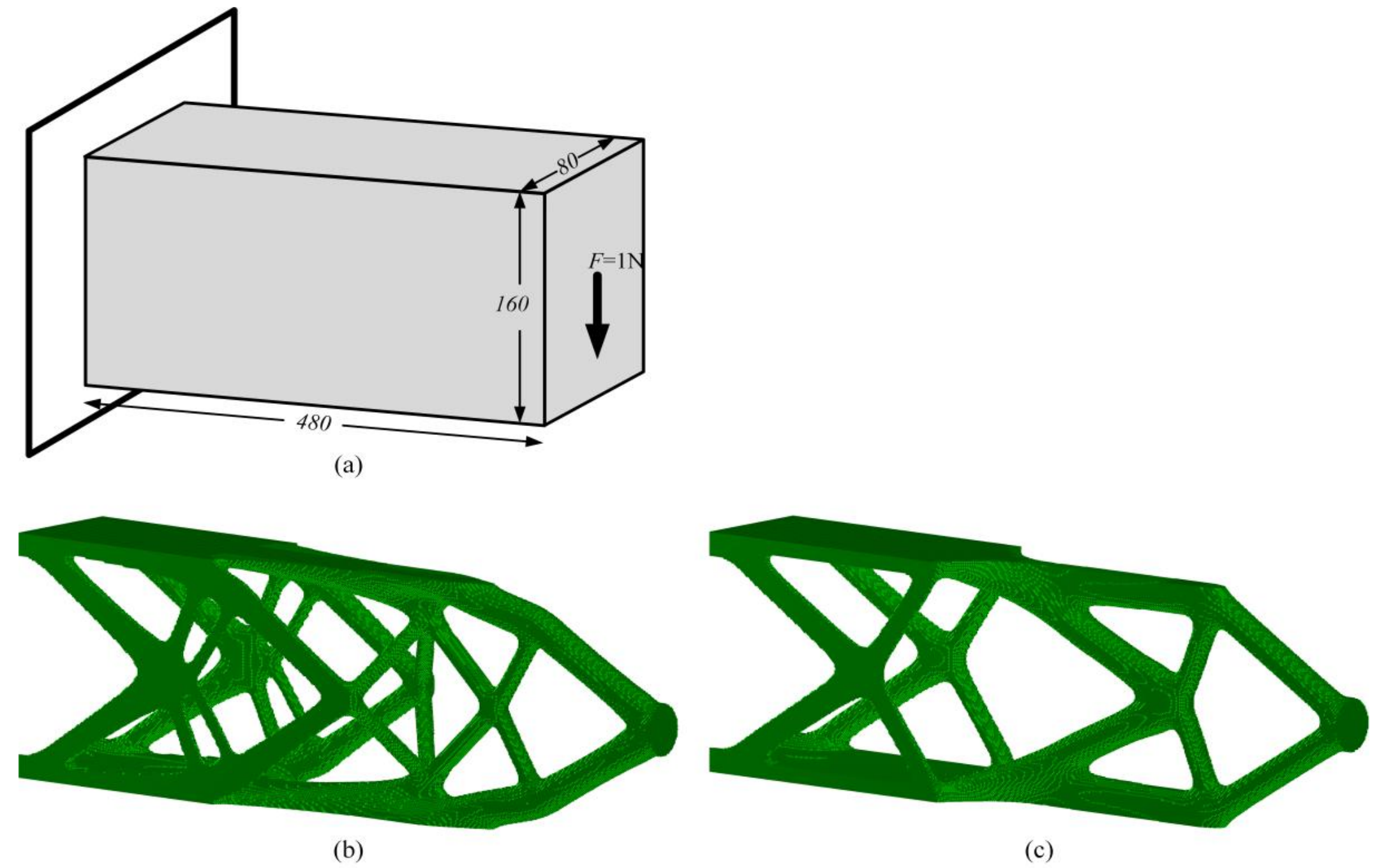

**Fig. 13** Design domain of the 3D cantilever beam and its topology solutions obtained from X-BESO method.

*6.4 Orthotropic box*

Now, we consider a rectangular domain (shown in Fig. 14(a)) with a traction load applied in the center bottom. The four corners of the design domain are constrained by plane joint. The model is discretized by 60×24×60 background meshes, and each element is partitioned with enriched nodes *En*=4. The total number of nodal design variables is 5633857. The maximum usable material volume fraction is set as 0.2. The optimization solution is given in Fig. 14(b), and total 18.7h is consumed to obtain the result. By comparison with section 6.3, it can be concluded that the coarser the finite mesh takes, the higher the computational efficiency is. But the optimization results are more accurate and the family of feasible structural domains is larger, if using finer mesh coupled with smaller numbers of enriched nodes.

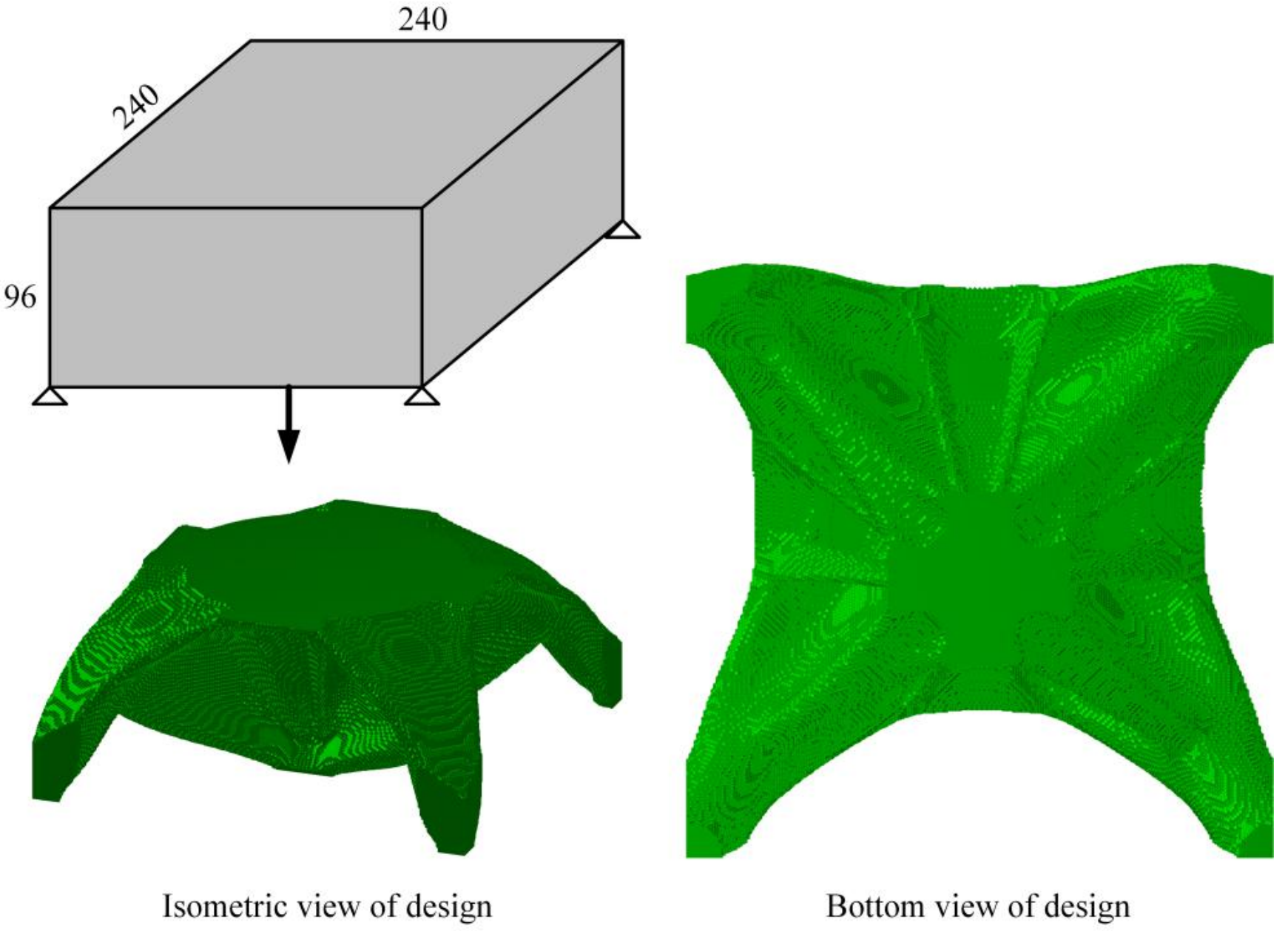

**Fig. 14** Design domain of the orthotropic box and its topology solutions obtained from X-BESO method.

*6.5 3D Bridge design*

For the last example, we consider the 3D bridge problem illustrated in Fig. 15(a). A uniform load is applied on a horizontal layer, and the design space is supported at four symmetric points at a distance of 32 units from two ends. The background finite mesh is 120×40×16, and the enriched nodes $En$=4 are defined. The 481×161×65 nodal design variables are allocated in the whole design domain. The filter radius is $r$=8.

Fig. 15(b) shows the optimized bridge with a volume fraction of 0.3. In the optimized bridge, catenary-like structure is produced to support the loading layer. The total number of nodal design variable for the bridge is 5033665, and about 7.1 hours are used to obtain the final result.

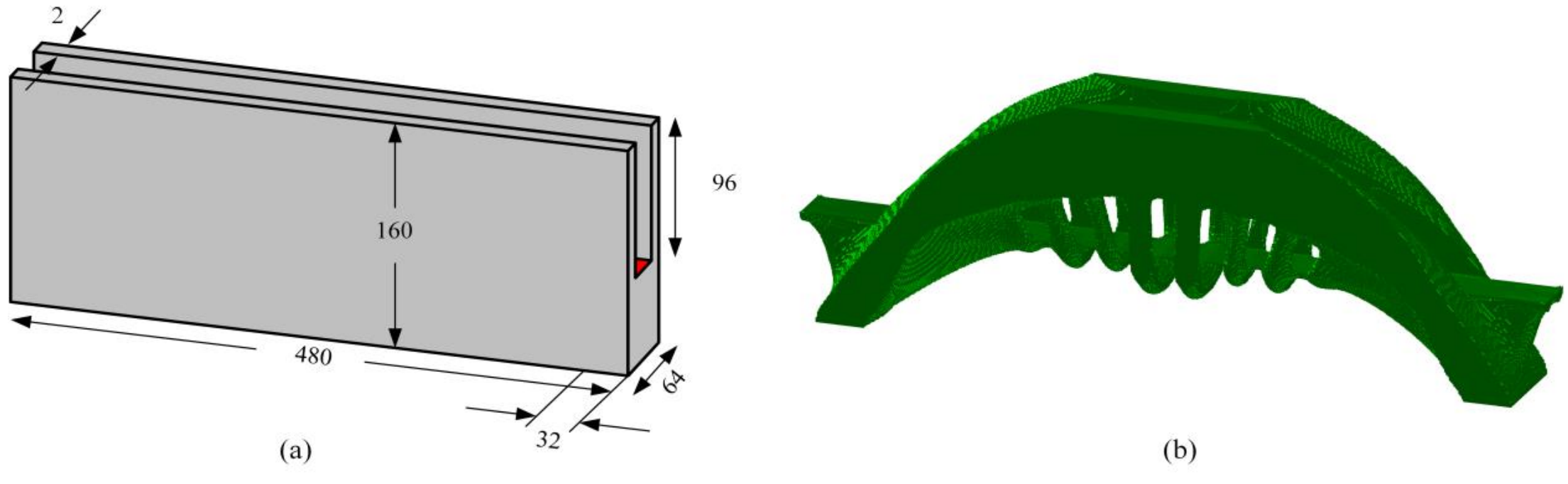

**Fig. 15** Design domain of the 3D bridge and its topology solutions obtained from X-BESO method.

## 7 Conclusions

In this work, a novel BESO method combined with XFEM is developed for large-scale or high-resolution topology optimization. The presented X-BESO method takes advantages of the ability of XFEM to accurately model material discontinuities within one element. Via allocating a set of enriched nodes, finite elements are uniformly divided into several sub-triangles or sub-tetrahedrons that carry individual material properties and shape functions. The Heaviside function is used to represent the material void discontinuity. Assembling the stiffness matrix of all the sub-parts into the global stiffness matrix, the sensitivity information for large number of nodal design variables could be obtained from a coarser finite mesh. Several examples involving millions of nodal design variables are illustrated to verify the applicability of the X-BESO method. The advantage of the presented method is that it could generate more accurate optimization solutions using tractable computational cost. At the same time, the X-BESO method alleviates the requirement on the computer's memory space. Moreover, the proposed triangulated partition for XFEM improves the quality of the boundary representation. Inspired from recently proposed Moving Morphable Components/Voids methods [32-34], the proposed X-BESO method would be further developed to consider the manufacturability of the design structures.

**Acknowledgements:**

This work was supported by the Key Program of National Natural Science Foundation of China (No. 11832009), the National Natural Science Foundation of China (No.11672104, 11902085), and the Chair Professor of Lotus Scholars Program in Hunan province (No.XJT2015408).